\documentclass[reprint, aps,pra,twocolumn,showpacs,amsmath,amssymb, floatfix]{revtex4-2}
\usepackage{graphicx}
\usepackage{dcolumn}
\usepackage{bm}
\usepackage{bbm}
\usepackage{amssymb}
\usepackage[table]{xcolor}
\usepackage{colortbl}
\usepackage{lipsum}
\usepackage{braket}
\usepackage{multirow}
\usepackage[colorlinks=true,linkcolor=blue, citecolor=red, urlcolor=magenta]{hyperref}
\usepackage{url}
\begin{document}

\title{Mathematics of effective $q$-plate in polarization optics}
\author{Mohammad Umar}
\email{aliphysics110@gmail.com}
\author{Paramasivam Senthilkumaran}
\email{psenthil@opc.iitd.ac.in}
\affiliation{Optics and Photonics Centre\\ Indian Institute of Technology Delhi\\ New Delhi 110016, INDIA}

\begin{abstract}
The $q$-plate is a spatially inhomogeneous SU(2) birefringent optical element that has garnered significant interest due to its ability to mediate the spin-orbit interaction of light and facilitate the generation of optical vortices. The $q$-plate features a spatially varying fast axis orientation defined by two parameters: the topological charge $q$ and the offset angle $\alpha_0$. The notion of an effective waveplate arises when multiple waveplates, whether homogeneous or inhomogeneous, are aligned coaxially such that, under specific constraints, the composite system emulates the behavior of a single effective waveplate. This work presents a comprehensive mathematical formalism for realizing an effective waveplate through a cascaded configuration of three $q$-plates, each chosen as either a quarter-wave $q$-plate, a half-wave $q$-plate, or a combination thereof. This yields a total of eight distinct configurations. Some configurations result in an effective waveplate exhibiting a constant retardance, whereas others allow continuous modulation of the effective retardance over the full range from $0$ to $2\pi$ through the systematic variation of the relative offset angles between the constituent $q$-plates. This feature enables holonomic polarization transformations on the higher-order Poincaré sphere, making the concept of the effective waveplate applicable to topological index spaces. Moreover, tunable effective retardance holds significant potential for applications involving structured light corresponding to the higher-order Poincar\'{e} sphere, particularly in scenarios demanding controlled spatial modulation of polarization states or dynamic tailoring of polarization topologies.
\end{abstract}

\maketitle

\section{Introduction}
A waveplate is a birefringent optical element that implements an SU(2) transformation on the polarization state of light and is characterized by two fundamental parameters, the retardance $\delta$ and the orientation of its fast axis $\alpha$ \cite{hecht2012optics, goldstein2003polarized, collett2003polarized}. The retardance denotes the phase shift induced between the orthogonal polarization components aligned along the fast and slow axes as the light propagates through the medium. Common examples include the quarter-wave plate (QWP) and half-wave plate (HWP), which exhibit retardance of $\pi/2$ and $\pi$ respectively. These are classified as homogeneous waveplates due to their spatially uniform parameters, constant retardance and fast axis orientation. In contrast, inhomogeneous waveplates exhibit spatially varying parameters, such as a position-dependent fast axis orientation or retardance. A notable example of such a device is the $q$-plate, wherein the fast axis orientation varies as a function of the azimuthal coordinates \cite{marrucci2006optical, marrucci2013q, rubano2019q, delaney2017arithmetic, machavariani2008spatially, kadiri2019wavelength, bansal2020use, cardano2012polarization, slussarenko2011tunable}.\\
\indent
A $q$-plate is also a birefringent SU(2) optical element where the fast axis orientation is azimuthally varying. Depending on the magnitude of the phase retardance $\delta$ introduced between orthogonal polarization components, the $q$-plate operates either as a quarter-wave $q$-plate ($q^{Q}$-plate) when $\delta = \pi/2$, or as a half-wave $q$-plate ($q^{H}$-plate) when $\delta = \pi$. Each $q$-plate is characterized by a topological charge $q$, which determines the number of complete $2\pi$ rotations that the fast axis orientation undergoes over one full azimuthal cycle in the transverse plane. A discussion on the $q$-plate is presented in Section \ref{c_01}. Owing to the spatial variation of the fast-axis orientation, the $q$-plate converts homogeneously polarized input light into a cylindrically polarized vector beam and induces a spin-orbit interaction, resulting in the transfer of orbital angular momentum (OAM) to the output beam \cite{marrucci2006optical}.\\
\indent
In polarization optics, topological index spaces have been recently introduced \cite{umar2025holonomically} to systematically describe polarization-structured electromagnetic fields and birefringent elements with engineered anisotropies exhibiting specific topologies. Each index space contains two types of members in the family, the topological sphere and the class of structured optical elements capable of performing holonomic polarization transformations \cite{umar2025holonomically, umar20252} on that sphere. In this index space, homogeneous waveplates are classified as members of the index zero space, while inhomogeneous waveplates belong to the higher index spaces.\\ 
\indent
The aim of this paper is to study the combination of three $q$-plates in general.. Combinations of homogeneous waveplates have been extensively studied, whereas investigations involving inhomogeneous waveplates remain largely unexplored, with the exceptions of the work presented in \cite{delaney2017arithmetic, kadiri2019wavelength}. The combinations of homogeneous waveplates used to find an SU(2) gadget for the Poincar\'{e} sphere (PS) were explored earlier. R. Simon and his collaborators published a series of three papers. In their initial work \cite{simon1989hamilton}, they demonstrated that a sequence of six waveplates (two HWPs and two QWPs) could perform any arbitrary SU(2) polarization transformation on the PS. This was optimized in subsequent work \cite{simon1989universal}, reducing the setup to four waveplates (two HWPs and two QWPs). In their final paper \cite{simon1990minimal}, they proposed a minimal configuration with just three waveplates, two QWPs and one HWP, arranged in any sequence (Q-H-Q, H-Q-Q, or Q-Q-H), achieving the full range of SU(2) operations. This configuration is now widely recognized as a \textit{universal} SU(2) gadget for polarization optics. The combination of two QWPs has also been studied as a polarization gadget, and it has been shown that this setup is sufficient to realize all polarization transformations on the PS \cite{reddy2014polarization}. Recently, an optical gadget has been proposed in which two HWPs are sandwiched between $q^{Q}$-plates to realize arbitrary polarization transformations on the HOPS \cite{bansal2025gadget}. Interestingly, this device incorporates optical elements belonging to different topological index spaces, an example of mixed index space. These studies are notable examples of research involving combinations of waveplates. \\
\indent
Moreover, in singular beams containing C-point and V-point singularity \cite{ruchi2020phase, senthilkumaran2024singularities}, polarization singularity sign inversion can be achieved using an HWP. \cite{pal2017polarization}. By placing the HWP appropriately with two $q^{H}$-plates, a new $q^{H}$-plate with the addition or subtraction of their topological charges can be achieved \cite{delaney2017arithmetic}.  Previous study has explored the concept of an effective waveplate and introduced the concept of a wavelength-adaptable effective waveplate \cite{kadiri2019wavelength}. Furthermore, the $q$-plate is of significant interest, and its numerous applications can be observed in different contexts \cite{marrucci2006optical, marrucci2013q, rubano2019q, delaney2017arithmetic, machavariani2008spatially, kadiri2019wavelength, bansal2020use, cardano2012polarization, slussarenko2011tunable, yao2011orbital, shu2017polarization, yao2023quantitative, gregg2015q, ji2016meta}. Recently a new type of $q$-plate is introduced where the retradnce of the waveplate is radially varying \cite{hakobyan2025q}.\\
\indent
Motivated by these advancements, we investigate the mathematical formulation of the effective waveplate realized through combinations of three coaxially aligned $q$-plates, each of which may be of the $q^{Q}$-type, the $q^{H}$-type, or a mixture thereof, yielding a total of eight distinct permutations. When multiple waveplates are arranged coaxially, the entire configuration can, under specific constraints, behave equivalently to a single waveplate. A canonical example is two QWPs arranged with aligned fast axes, effectively behaving as a HWP. The essential constraint governing such effective behavior is the alignment of the fast axis orientation. In the present study, we demonstrate that certain configurations exhibit tunable retardance ranging continuously from $0$ to $2\pi$ and $\pi/2$ to $3\pi/2$. This tunability is governed by the relative offset angles of the constituent $q$-plates. To ensure the completeness of this draft, it is essential to include a discussion on the PS (a member of topological index zero space), the higher-order Poincaré sphere (HOPS) (member of higher order topological index spaces) and the action of a the homgeneous waveplate and the $q$-plate on the PS and HOPS, respectively.\\
\indent
The structure of this paper is organized as follows. Section \ref{section02} introduces the underlying topological constructs together with the associated polarization elements, which are essential for establishing the completeness of the present study. In particular, this section provides a detailed discussion of the PS, HOPS and the SU(2) action of the homogeneous waveplate and inhomogeneous $q$-plate. Section \ref{section03} focuses on the concept of the effective waveplate, providing its mathematical formulation. Combinations of two and three $q$-plates are studied under the concept of the effective waveplate. Section~\ref{section04} presents the concluding remarks, summarizing the outcomes.
\section{Polarization Formalism}
\label{section02}
\subsection{The (higher-order) Poincar\'{e} sphere}
\begin{figure*}[t]
\centering
\includegraphics[width=0.9\linewidth]{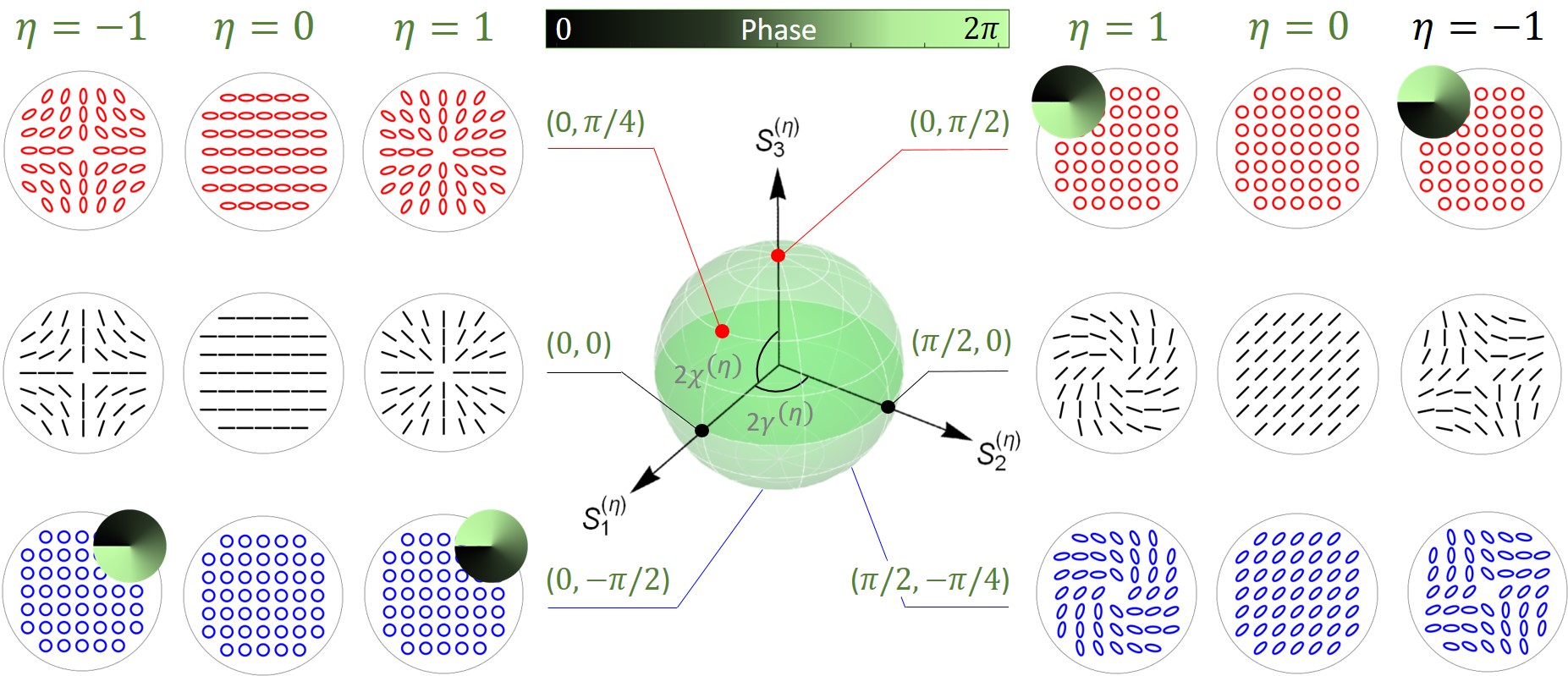}
\caption{(Color online). The geometry of the standard PS ($\eta=0$) and HOPS ($\eta=-1$ and $+1$), along with their corresponding polarization distributions depicted as individual points on the surface, are shown. Here, $2\gamma^{(\eta)}$ and $2\chi^{(\eta)}$ represent the longitude and latitude coordinates, respectively. The red and blue colors correspond to right- and left-handed polarization. Additionally, the vortex phase, embedded with RCP and LCP for $\eta=-1$ and $+1$, is also shown.}
\label{q01}
\end{figure*}
Polarization optics utilizes different orders of topological spheres as models to describe uniform as well as spatially varying polarization states. Each sphere encapsulates a distinct polarization texture within the optical field, characterized by its associated topological indices that quantify the underlying singularities and symmetries. The canonical example is the Poincar\'{e} sphere, an $\textbf{S}^{2}$ sphere, introduced by Henri Poincar\'{e} in 1892 \cite{poincare1954theorie}, provides a bijective geometric representation of the state of polarization (SOP). It maps all polarization states onto its surface using the Stokes parameters (SPs) $S_{1}$, $S_{2}$ and $S_{3}$ as Cartesian coordinates, subject to the condition $S_{0}^{2} = S_{1}^{2} + S_{2}^{2} + S_{3}^{2}$ for fully polarized light. For the partially polarized light $S_{0}^{2} > S_{1}^{2} + S_{2}^{2} + S_{3}^{2}$. Each point on the sphere corresponds bijectively to a distinct SOP. The equator of the PS is populated by linearly polarized light, while the poles are associated with circularly polarized light. The rest of the surface corresponds to elliptically polarized light. The northern and southern hemispheres are used to distinguish right-handed and left-handed polarization. The latitude and longitude coordinates on the PS are defined by $2\gamma = \tan^{-1}(S_{2}/S_{1})$ and $2\chi = \sin^{-1}(S_{3}/S_{0})$, where $\gamma$ and $\chi$ represent the azimuth and ellipticity of the polarization ellipse, respectively.\\
\indent
Despite the widely recognized geometric elegance of the PS, it falls short when representing higher-order solutions of Maxwell’s equations, which admits spatially inhomogeneous beams. While homogeneous polarization states elegantly correspond to single points on the PS, structured beams occupy extended, interconnected areas, revealing the limitations of the PS. Addressing this, G. Milione et al. proposed the higher-order Poincar\'{e} sphere (HOPS) \cite{milione2011higher}. The elegance of HOPS lies in its ability to represent spatially inhomogeneous beams of constant ellipticity succinctly as distinct points on its surface, thereby extending the concept of standard PS. The HOPS, like the standard PS, is also an $\textbf{S}^{2}$ sphere where the higher-order SPs serve as Cartesian coordinates. While the PS functions as a Bloch sphere whose basis is defined solely by spin angular momentum (SAM) polarization states, the HOPS extends this framework by incorporating both SAM and orbital angular momentum (OAM).\\
\indent
The HOPS beam is expressed as
\begin{equation}
    |\psi_{\ell}\rangle = \psi_{R}\ket{R_\ell}+\psi_{L}\ket{L_\ell},
    \label{eqn_psi}
\end{equation}
where $|R_{\ell}\rangle$ and $|L_{\ell}\rangle$ are the orthogonal basis states embedded with optical vortex of topological charge of $-\ell$ and $\ell$ respectively. The Poincar\'e-Hopf (PH) index of the HOPS beam is half the difference between the topological charges of the vortex corresponding to LCP and RCP light, i.e. $\eta=\frac{1}{2}(\ell-(-\ell))=\ell$. The term $\psi_{R}$ and $\psi_{L}$ are the complex amplitudes expressed as 
\begin{align}
    \psi_{R} &= \bra{R_\ell}\psi_{\ell}\rangle = \frac{1}{\sqrt{2}}\left[\cos\chi^{(\eta)}+\sin\chi^{(\eta)} \right]e^{-i\gamma^{(\eta)}}, \\
    \psi_{L} &= \bra{L_\ell}\psi_{\ell}\rangle = \frac{1}{\sqrt{2}}\left[\cos\chi^{(\eta)}-\sin\chi^{(\eta)} \right]e^{i\gamma^{(\eta)}}.
\end{align}
\noindent
Here, $2\gamma^{(\eta)}$ and $2\chi^{(\eta)}$ represent the longitude and latitude coordinates on the HOPS, respectively. For $\eta=0$, the HOPS reduces to the standard PS, with the longitude and latitude coordinates $2\gamma^{(0)}$ and $2\chi^{(0)}$, respectively. For $\eta \neq 0$, $2\gamma^{(\eta)}$ and $2\chi^{(\eta)}$ are related to the Pancharatnam phase \cite{pancharatnam1956generalized} and ellipticity, respectively \cite{bansal2023stokes}. On the HOPS, the equator is populated by the cylindrical vector beams, including radially, spirally and azimuthally polarized beams, and this occurs when $|\psi_{R}| = |\psi_{L}|$ while non-equatorial points correspond to cases when $|\psi_{R}| \neq |\psi_{L}|$. The polarization distribution corresponding to the HOPS of $\eta=-1$, $0$ and $+1$ is presented in Fig. \ref{q01}. The SPs corresponding to the HOPS beam is given by $S_{0}^{(\eta)} = |\psi_R|^{2} + |\psi_L|^{2}$, $S_{1}^{(\eta)} = 2\texttt{Re}[\psi_{R} \psi_{L}^{*}]$, $S_{2}^{(\eta)} = 2\texttt{Im}[\psi_{R} \psi_{L}^{*}]$ and $S_{3}^{(\eta)} = |\psi_R|^{2} - |\psi_L|^{2}$.
Here, the parameters $S_{1}^{(\eta)}$, $S_{2}^{(\eta)}$, and $S_{3}^{(\eta)}$ serve as the Cartesian coordinates in the construction of HOPS, with the relation $(S_{0}^{(\eta)})^2 = (S_{1}^{(\eta)})^2 + (S_{2}^{(\eta)})^2 + (S_{3}^{(\eta)})^2$. In terms of these SPs, the coordinates $2\gamma^{(\eta)}$ and $2\chi^{(\eta)}$ are given by $2\gamma^{(\eta)} = \tan^{-1}(S_{2}^{(\eta)} / S_{1}^{(\eta)})$ and $2\chi^{(\eta)} = \sin^{-1}(S_{3}^{(\eta)} / S_{0}^{(\eta)})$, respectively.\\
\indent
The emergence of the HOPS has generated considerable attention within the optics community. A wide range of investigations has been devoted to different aspects of HOPS beams, encompassing their generation \cite{naidoo2016controlled, chen2014generation, ji2023controlled, liu2024generation, yao2022generation}, polarization evolution via metasurfaces \cite{liu2014realization} and photonic Dirac points \cite{xu2021polarization}, as well as their detection \cite{yao2022generation, yao2023quantitative}. Advances have also been reported in Stokes polarimetry \cite{bansal2023stokes} and polarimetry employing all-Silicon metadevices \cite{yang2021all}. Further studies have explored the spatiospectral higher-order topological sphere \cite{fickler2024higher} and the interaction of atoms with HOPS optical vortex modes \cite{bougouffa2025optical}. The tight focusing of higher-order Poincar\'{e} beams has also been analyzed \cite{pal2024tight}. More recently, a novel polarimetry approach leveraging metasurface photonics for polarization detection specifically tailored to HOPS has been demonstrated \cite{yang2025metasurface}.
\subsection{SU(2) action of a homogeneous waveplate}
The homogeneous waveplate is defined by two parameters, its retardance $\delta$ and the orientation $\alpha$ of its fast axis relative to a chosen reference axis. In Jones formalism, this waveplate is represented by an SU(2) matrix as \cite{collett2003polarized}
\begin{equation}
M(\delta, \alpha)= 
\begin{bmatrix}
        \cos \frac{\delta}{2} + i\sin\frac{\delta}{2} \cos 2 \alpha & i\sin\frac{\delta}{2} \sin 2 \alpha \\[14pt]
        i\sin\frac{\delta}{2} \sin 2 \alpha & \cos \frac{\delta}{2} - i\sin\frac{\delta}{2} \cos 2 \alpha
\end{bmatrix}.
\label{jones_matrix}
\end{equation}
This matrix is symmetric, satisfying the condition $M(\delta, \alpha) = [M(\delta, \alpha)]^{T}$. The Jones matrices corresponding to the QWP and HWP satisfy the eighth and fourth roots of the identity element, respectively, expressed mathematically as 
\begin{equation}
[M(\pi/2, \alpha)]^{8} = \textbf{I} \quad \text{and} \quad [M(\pi, \alpha)]^{4} =  \textbf{I}.
\label{equation_05}
\end{equation}
The diagonal elements are complex conjugates, i.e., $M_{11}(\delta, \alpha) = [M_{22}(\delta, \alpha)]^{*}$, while the off-diagonal elements are identical i.e., $M_{12}(\delta, \alpha) = M_{21}(\delta, \alpha)$, and are purely imaginary. These are the SU(2) matrices $(\texttt{det}[M(\delta, \alpha)]=1$ and $M(\delta, \alpha)[M(\delta, \alpha)]^{\dagger}=\mathbb{I})$, enabling norm-preserving polarization transformations. If the Jones matrix of a waveplate is given then we can extract the information about the retardance $\delta$ and fast axis orientation $\alpha$ as
\begin{align}
\delta &= 2\cos^{-1}\Bigg[\frac{1}{2}\texttt{Tr}[M(\delta, \alpha)]\Bigg] \label{delta}, \\
\alpha &= \frac{1}{2}\tan^{-1}\Bigg[\frac{\texttt{Im}[M_{12}(\delta, \alpha)]}{\texttt{Im}[M_{11}(\delta, \alpha)]}\Bigg].
\label{alpha}
\end{align}
respectively. Next, revisiting the SU(2) Jones matrix from Eq. (\ref{jones_matrix}), it can be expressed in a notational form as
\begin{equation}
M(\delta, \alpha)= 
\begin{bmatrix}
        \mathbbmss{A}+i\mathbbmss{B} & i\mathbbmss{D} \\[14pt]
        i\mathbbmss{D} & \mathbbmss{A}-i\mathbbmss{B}
\end{bmatrix} \in \text{SU}(2),
\label{matrix02}
\end{equation}
where, $\mathbbmss{A}$, $\mathbbmss{B}$ and $\mathbbmss{D}$ are the function $f(\delta, \alpha)$. It is important to mention again that off-diagonal elements are the same and only have imaginary components. This is a \textit{constraint} present in the Jones matrix of a waveplate and we will build the concept of the effective waveplate based on this constraint, which will come later.

The SU(2) matrix in exponentially parameterized form, with a rotation axis $\textbf{k}$ and a rotation angle $\delta$, is given by \cite{bettegowda2017prescription, damask2004polarization}
\begin{equation}
\begin{aligned}
    M (\delta, \mathbf{k}) &= e^{i\frac{\delta}{2}(\textbf{k}\cdot\sigma)} 
       = \cos{\frac{\delta}{2}}\textbf{I}+i\sin\frac{\delta}{2}(\textbf{k}\cdot\sigma),
\end{aligned}
\label{su2_matrix}
\end{equation}
where, $\textbf{I}$ is the identity matrix and $\sigma$ is the Pauli spin matrix whose components are $\sigma_{j}$, $j=x,y,z$ and given by
\begin{equation}
\sigma_x = \begin{bmatrix} 0 & 1 \\ 1 & 0 \end{bmatrix}, \quad
\sigma_y = \begin{bmatrix} 0 & -i \\ i & 0 \end{bmatrix}, \quad
\sigma_z = \begin{bmatrix} 1 & 0 \\ 0 & -1 \end{bmatrix}.
\end{equation}
Parameterizing the rotation axis as $\mathbf{k}=(\cos{2\alpha},\sin{2\alpha},0)^T$, Eq. (\ref{su2_matrix}) becomes the Jones matrix of a homogeneous waveplate, as expressed in Eq. (\ref{jones_matrix}). Consequently, the retardance $\delta$ and the fast-axis angle $\alpha$ in the SU(2) waveplate space correspond directly to the rotation angle and the orientation of the rotation axis, respectively, in the SO(3) PS space. SO(3) is related to SU(2) through a two-to-one surjective homomorphism, with SU(2) serving as its double cover. Since the fast-axis orientation in the waveplate is restricted to a two-dimensional plane (the transverse plane of the waveplate), the corresponding rotation axis in the SO(3) PS space is parameterized as $\mathbf{k}=(k_{x}=\cos2\alpha, k_{y}=\sin2\alpha, k_{z}=0)^{T}$, lying within a two-dimensional plane, the equatorial plane. In this plane, the rotation axis subtends an angle of $2\alpha$ with the $S_{1}$-axis of the PS \cite{kumar2011polarization}. Thus, the action of the waveplate on a homogeneously polarized light beam is interpreted as a rotation by an angle (equal to $\delta$) about the rotation axis, which makes an angle of $2\alpha$ with the $S_{1}$-axis, when viewed from $S_{1}$ side. For a detailed visualization of how polarization states evolve as a rotation on the PS under the action of a waveplate, refer to \cite{kumar2011polarization}. Here, the discussion pertains to the topological index-zero space. We now extend the discussion to the index-$\eta$ space, whose members include the $q$-plate and the HOPS.
\begin{figure}[t]
\centering
\includegraphics[width=0.9\linewidth]{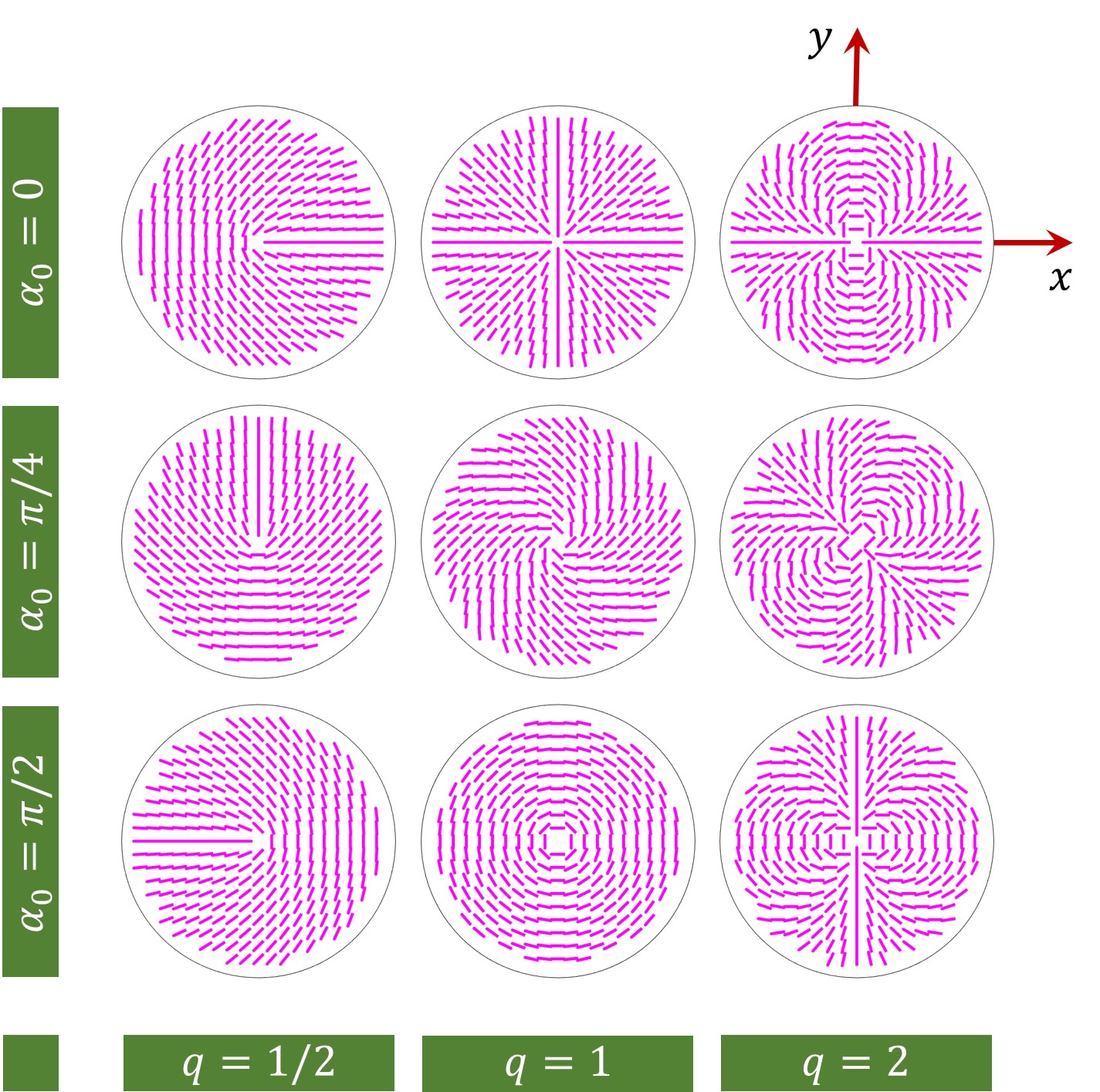}
\caption{(Color online). Geometry of the $q$-plate of topological charge $q=1/2$ (first column), $q=1$ (second column) and $q=2$ (third column) with different offset angles, $\alpha_{0} = 0$ (first row), $\alpha_{0} = \pi/4$ (second row) and $\alpha_{0}=\pi/2$ (third row). The offset angle refers to the orientation of the fast axis relative to a specified reference axis, which in this case is the $x$-axis.}
\label{abc}
\end{figure}
\subsection{The (inhomogeneous) \textit{q}-plate}
\label{c_01}
The homogeneous waveplates, QWP and HWP, are members of the SU(2) group and have uniform retardance and fast axis orientation throughout their transverse plane. However, if any of these two parameters is not uniform throughout the plane of the plate, the plate is no longer homogeneous and it becomes inhomogeneous. The $q$-plate \cite{marrucci2006optical, marrucci2013q, rubano2019q, delaney2017arithmetic, machavariani2008spatially, kadiri2019wavelength, bansal2020use, cardano2012polarization, slussarenko2011tunable} is a well-known example of an inhomogeneous waveplate and is also a member of the SU(2) group. In the $q$-plate (as shown in Fig. \ref{abc}), the fast axis orientation varies azimuthally in its transverse plane, while its retardance remains fixed. The fast axis orientation of the $q$-plate is given by
\begin{equation}
    \alpha(\phi) = q\phi + \alpha_{0}.
\end{equation}
Here, $q$ is the topological charge, which quantifies the amount of rotation of the fast axis in a complete $2\pi$ azimuthal rotation, and $\alpha_{0}$ represents the offset angle, which is the orientation of the fast axis with respect to a reference axis. Depending on the retardance of the $q$-plate (or $q^{\delta}$-plate), it can be classified as a quarter-wave $q$-plate ($q^{Q}$-plate) or a half-wave $q$-plate ($q^{H}$-plate), where the retardance values are $\pi/2$ and $\pi$, respectively. In Eq. (\ref{jones_matrix}), replacing $\alpha \rightarrow \alpha(\phi)$ gives the Jones matrix of the $q$-plate with retardance $\delta$, which is symmetric and respects SU(2) features. $q^{Q}$-plate and $q^{H}$-plate also respects the Eq. (\ref{equation_05}).
\subsection{SU(2) action of a \textit{q}-plate}
The HOPS beam and the $q$-plate share a common topological structure, as both are characterized by azimuthal variation in one of their defining parameters. Leveraging this shared topology, we have demonstrated SU(2) polarization evolution on the HOPS using a $q$-plate \cite{umar20252, umar2025holonomically}. For such evolution to occur, a holonomy condition must be satisfied, specifically, to perform a holonomic transformation on a HOPS of order $\eta$, the $q$-plate must possess a topological charge of $q$, which should be equal to the order $\eta$ of the HOPS.
\begin{equation}
\label{holonomy}
\underbrace{\ell=\eta=q}_{\text{Holonomy condition}}
\end{equation}
\indent
Under the holonomy condition, the SU(2) action of the $q$-plate on the HOPS beam manifests as an SO(3) rotation on the HOPS \cite{umar2025holonomically, umar20252}. Furthermore, it has been shown that a single \textit{global} SO(3) rotation on the HOPS corresponds to multiple \textit{local} SO(3) rotations on the standard PS \cite{umar20252}. Consequently, two types of rotations are involved: one requiring a global rotation axis and the other involving multiple local rotation axes. The fast axis orientation of the $q$-plate is given by $\alpha(\phi) = q\phi + \alpha_0$, comprising a $\phi$-dependent term $q\phi$ and a $\phi$-independent term $\alpha_0$. The offset angle $\alpha_0$ determines the global rotation axis for the rotation on the HOPS, while the $q\phi$ term governs the multiple local rotation axes for the rotations on the the standard PS. In both spheres, the rotation axis is always lying in the equatorial plane.
\section{Theory}
\label{section03}
\subsection{Effective waveplate}
Multiple waveplates used in a sequence, under some specific conditions, can emulate the performance of a single waveplate, which is termed as an effective waveplate.  This effective waveplate is characterized by effective retardance $\delta_{e}$ and effective fast axis orientation $\alpha_{e}$. Consider $n$ number of waveplates having retardances $\delta_{1}$, $\delta_{2}$, $\delta_{3}$, \ldots, $\delta_{n}$ and fast axis orientations $\alpha_{1}$, $\alpha_{2}$, $\alpha_{3}$, \ldots, $\alpha_{n}$ respectively are placed coaxially. By using the composition property of the matrix the whole setup mathematically can be expressed as
\begin{multline}
\mathbb{M} = M(\delta_{n}, \alpha_{n}) \cdot M(\delta_{n-1}, \alpha_{n-1})\cdot . . . . . \\
. . . . . \cdot M(\delta_{3}, \alpha_{3})\cdot M(\delta_{2}, \alpha_{2}) \cdot M(\delta_{1}, \alpha_{1}).
\label{composition_matrix}
\end{multline}
After the multiplication of these Jones matrices, the resulting matrix $\mathbb{M}$ may not necessarily conform to the form presented in Eq. (\ref{matrix02}), instead, it can also yield a matrix whose off-diagonal elements include real components. In the matrix form Eq. (\ref{composition_matrix}) is expressed as
\begin{equation}
\mathbb{M}= 
\begin{bmatrix}
        \mathbbmss{A}+i\mathbbmss{B} & \mathbbmss{C}+i\mathbbmss{D} \\[14pt]
        -\mathbbmss{C}+i\mathbbmss{D} & \mathbbmss{A}-i\mathbbmss{B}
\end{bmatrix}.
\label{matrix04}
\end{equation}
Thus, the resultant matrix corresponding to a combination of waveplates differs from that of a single waveplate due to the presence of the real component $\mathbbmss{C}$ in the off-diagonal elements. To achieve a behavior equivalent to a single effective waveplate, it is essential that the real component $\mathbbmss{C}$ vanishes. This condition represents the specific \textit{constraint} for a sequence of waveplates to function effectively as a single waveplate. In this matrix, the elements $\mathbbmss{A}$, $\mathbbmss{B}$, $\mathbbmss{C}$ and $\mathbbmss{D}$ are functions of the form 
\begin{equation}
f(\delta_{1},\delta_{2},\delta_{3},\ldots,\delta_{n}, \alpha_{1},\alpha_{2},\alpha_{3},\ldots,\alpha_{n}).
\end{equation} 
Therefore, by an appropriate choice of these waveplate parameters, one can arrange the system such that the real term $\mathbbmss{C}$ vanishes. Therefore, for the $\mathbbmss{C}=0$ case, $\mathbb{M}=M(\delta_{e}, \alpha_{e})$.
\subsection{Combination of two $q$-plates}
\begin{figure}[t]
\centering
\includegraphics[width=0.8\linewidth]{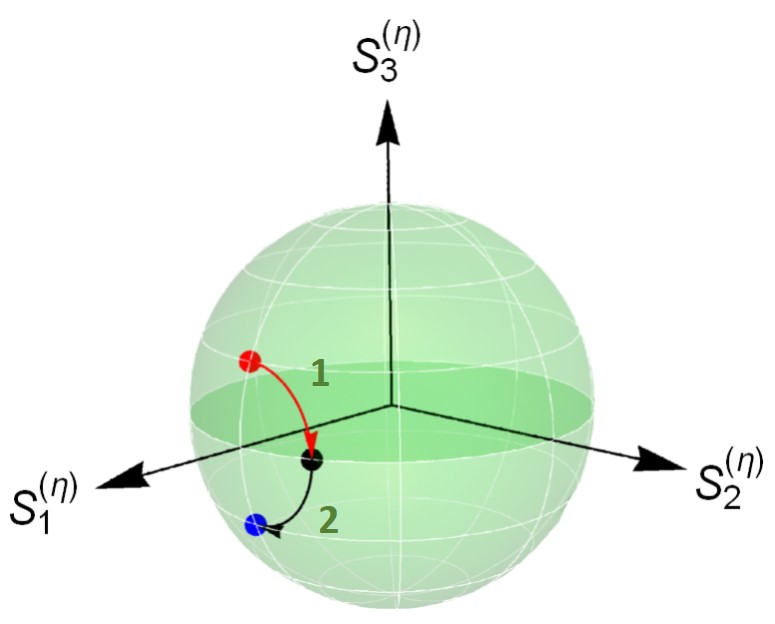}
\caption{(Color online). Depiction of the SOP evolution on the Poincaré sphere of order $\eta$, starting from the point $(2\gamma^{(0)}, 2\chi^{(0)}) = (0, \pi/8)$ (red point), as it passes through two sequential $q^{Q}$-plates of charge $q=\eta$ with identical fast-axis textures. The final output SOP is represented by the blue point at $(2\gamma^{(0)}, 2\chi^{(0)}) = (0, -\pi/8)$. This figure is valid for any topological index space.}
\label{trajectory_figure}
\end{figure}

Consider the combination of two $q$-plates with retardances $\delta_{1}$ and $\delta_{2}$, and fast axis orientations given by $\alpha_{1}(\phi) = q_{1}\phi + \alpha_{01}$ and $\alpha_{2}(\phi) = q_{2}\phi + \alpha_{02}$. Here, $q_{1}$ and $q_{2}$ represent the topological charges, while $\alpha_{01}$ and $\alpha_{02}$ are the offset angles of the respective plates. This arrangement can be described as
\begin{equation}
\begin{aligned}
\mathbb{M} & = M(\delta_{2}, \alpha_{2}) \cdot M(\delta_{1}, \alpha_{1}) \\[8pt]
        & =
\begin{bmatrix}
\mathbbmss{A}_{q}+i\mathbbmss{B}_{q} & \mathbbmss{C}_{q}+i\mathbbmss{D}_{q} \\[14pt]
-\mathbbmss{C}_{q}+i\mathbbmss{D}_{q} & \mathbbmss{A}_{q}-i\mathbbmss{B}_{q}
\end{bmatrix} \in \text{SU}(2).
\end{aligned}
\label{matrix05_06}
\end{equation}
where, $\mathbbmss{A}_{q}$, $\mathbbmss{B}_{q}$, $\mathbbmss{C}_{q}$ and $\mathbbmss{D}_{q}$ are expressed as
\begin{widetext}
\begin{align}
    \mathbbmss{A}_{q} &= \cos\frac{\delta_{1}}{2}\cos\frac{\delta_{2}}{2}
             - \sin\frac{\delta_{1}}{2}\sin\frac{\delta_{2}}{2}\cos2(\alpha_{1}(\phi)-\alpha_{2}(\phi))
\end{align}
\end{widetext}
\begin{widetext}
\begin{align}
    \mathbbmss{B}_{q} &= \sin\frac{\delta_{1}}{2}\cos\frac{\delta_{2}}{2}\cos2\alpha_{1}(\phi)
             + \cos\frac{\delta_{1}}{2}\sin\frac{\delta_{2}}{2}\cos2\alpha_{2}(\phi) \\[0.5ex]
    \mathbbmss{C}_{q} &= -\sin\frac{\delta_{1}}{2}\sin\frac{\delta_{2}}{2}
                \sin2(\alpha_{1}(\phi)-\alpha_{2}(\phi)) \\[0.5ex]
    \mathbbmss{D}_{q} &= \sin\frac{\delta_{1}}{2}\cos\frac{\delta_{2}}{2}\sin2\alpha_{1}(\phi)
              + \cos\frac{\delta_{1}}{2}\sin\frac{\delta_{2}}{2}\sin2\alpha_{2}(\phi)
\end{align}
\end{widetext}
It is evident that $\mathbbmss{C}_{q}$ is not zero, and to make this term zero, the condition $\alpha_{1}(\phi) - \alpha_{2}(\phi)=0$ should be respected. This condition is respected when $\alpha_{1}(\phi) - \alpha_{2}(\phi)=m\frac{\pi}{2}$, where $m$ is an integer. Applying this condition and calculating the effective retardance $\delta_{e}$ and effective fast axis orientation $\alpha_{e}$ by using Eq. (\ref{delta}) and (\ref{alpha}) respectively, we get
\begin{equation}
\delta_{e} =
\left\{
\begin{array}{ll}
  2\cos^{-1}\!\left[\cos\!\left(\dfrac{\delta_{1}+\delta_{2}}{2}\right)\right], 
  & m = 0, \pm 2, \pm 4, \cdots \\[4ex]
  2\cos^{-1}\!\left[\cos\!\left(\dfrac{\delta_{1}-\delta_{2}}{2}\right)\right], 
  & m = \pm 1, \pm 3, \cdots
\end{array}
\right.
\label{twoqdeltae}
\end{equation}
\begin{equation}
    \quad \alpha_{e} = \alpha_{2}(\phi) = q_{2}\phi + \alpha_{02}, \quad m =0, \pm 1, \pm 2, \cdots
\end{equation}
\indent
Now, consider the simple case of two QWPs which belong to zero index space ($q = 0$, $\delta_{1} = \pi/2$ and $\delta_{2} = \pi/2$) with their fast axes aligned, which corresponds to $m = 0$. In this case, $\delta_{e} = \pi$ and $\alpha_{e} = \alpha_{02}$. This shows that the combination of two QWPs with the aligned fast axis orientation effectively functions as a single HWP. This is a well-established, fundamental, and canonical example. This feature can also be seen geometrically on the PS. Consider the geometry depicted in Fig. \ref{trajectory_figure} for the case $\eta=0$. The red point represents a polarization state with coordinates $(2\gamma^{(0)}, 2\chi^{(0)}) = (0, \pi/8)$. Upon passing through a QWP with $\alpha = 0$, the output is at the black point with coordinates $(2\gamma^{(0)}, 2\chi^{(0)}) = (\pi/8, 0)$ and this transformation is shown by trajectory $1$. Further, this output is 
subsequently passed through another QWP with $\alpha = 0$, resulting in the polarization state (blue point) at $(2\gamma^{(0)}, 2\chi^{(0)}) = (0, -\pi/8)$, as illustrated by trajectory $2$. These trajectories showing the rotation about the $S^{(0)}_{1}$-axis. The transformation from the red to the blue point can also be achieved, via the same trajectory (trajectory 1 $+$ trajectory 2), by an HWP with $\alpha=0$. This shows that, under specific condition, two QWPs can effectively function as a single HWP. Furthermore, if two QWPs are oriented with their fast axes perpendicular to each other ($m = \pm 1$ case), then from Eq. (\ref{twoqdeltae}) it follows that $\delta_{e} = 0$, which corresponds to a trajectory represented by an arc of length zero in Fig. \ref{trajectory_figure}. Hence, in this configuration, the combined effect of the two QWPs cancels out.\\
\indent
Next, consider the combination of two $q^{Q}$-plates belonging to non-zero index space ($\delta_1 = \pi/2$, $\delta_2 = \pi/2$) with the same fast-axis textures, i.e., $m=0$. Accordingly, the effective retardance is $\delta_e = \pi$ and the effective fast axis orientation is given by $\alpha_e = q_2 \phi + \alpha_{02}$. This shows that the combination of two $q^{Q}$-plates behaves effectively like a single $q^{H}$-plate, as depicted again in Fig. \ref{trajectory_figure}. The key condition is that the fast axis orientations of both $q$-plates must be the same, which ensures that $\mathbbmss{C}_q = 0$. Furthermore, if we consider two $q^{Q}$-plates with $m = \pm 1$, then $\delta_{e} = 0$, and the effect of this configuration is null. However, this null effect applies only to beams within the same topological index space and light beams from a different index space will still experience an effect, like non-holonomic polarization transformations.\\
\indent
The realization of the effective waveplate for the combination of two waveplates (whether homogeneous or inhomogeneous) is straightforward. However, this becomes more complex when considering three plates, and that is the focus of this paper. In the following section, we provide the mathematical formulation necessary to describe the effective waveplate arising from the combination of three $q$-plates. Each of these plates can be selected as either a $q^{Q}$-plate, a $q^{H}$-plate or a mixture thereof, leading to a total of eight possible distinct configurations. Furthermore, by setting $q = 0$ within this framework, the same formulation naturally reduces to the case of three homogeneous waveplates. This unified treatment thus allows us to address both inhomogeneous and homogeneous situations under a common approach.
\subsection{Combination of three \textit{q}-plates}
\begin{figure}[t]
\centering
\includegraphics[width=1\linewidth]{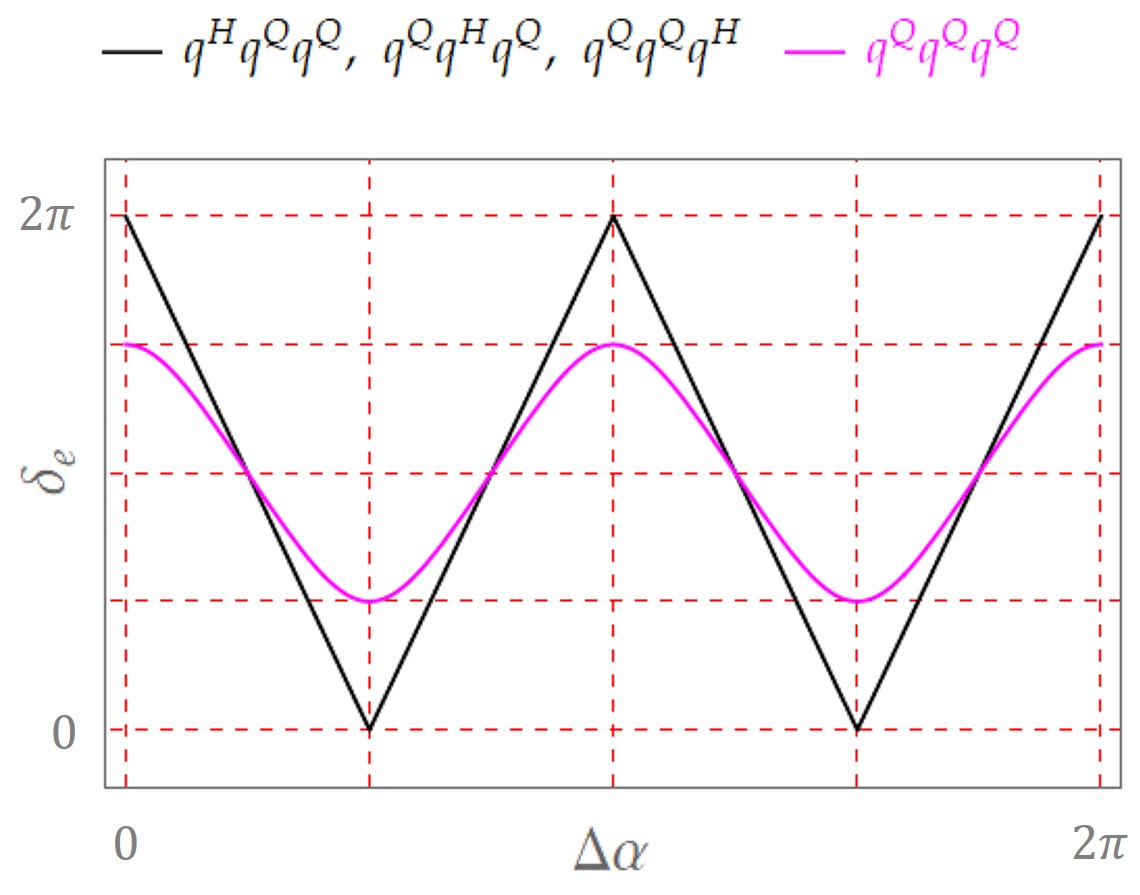}
\caption{(Color online). The variation of the effective retardance $\delta_{e}$ and the relative offset angle orientation $\Delta\alpha$ (black curve) for the $\mathcal{E}_1$, $\mathcal{F}_1$, and $\mathcal{G}_1$ cases (corresponding to the $q^{Q}q^{Q}q^{H}$, $q^{Q}q^{H}q^{Q}$, and $q^{H}q^{Q}q^{Q}$ configurations, respectively) is shown. In these cases, $\Delta \alpha = \alpha_{Q}-\alpha_{H}$. The magenta curve illustrates the same for the $\mathcal{H}_1$ case (the $q^{Q}q^{Q}q^{Q}$ configuration), where $\Delta \alpha = \alpha_{02}-\alpha_{03}$ (Eq. (\ref{qqq})).}
\label{plot}
\end{figure}

Consider three $q$-plates having retardance $\delta_{1}$, $\delta_{2}$ and $\delta_{3}$ and fast axis orientation $\alpha_{1}(\phi)=q_{1}\phi+\alpha_{01}$, $\alpha_{2}(\phi)=q_{2}\phi+\alpha_{02}$ and $\alpha_{3}(\phi)=q_{3}\phi+\alpha_{03}$ are placed coaxially. The overall effect of this arrangement can be described by using the matrix composition property, as shown below
\begin{equation}
\mathbb{M}  = M(\delta_{3}, \alpha_{3}) \cdot M(\delta_{2}, \alpha_{2}) \cdot M(\delta_{1}, \alpha_{1}) \\[8pt]
\end{equation}
The matrix form of this equation resembles the matrix presented in Eq. (\ref{matrix04}), with the corresponding terms given as
\begin{multline}
    \mathbbmss{A}_{q} = \cos\frac{\delta_{1}}{2}\cos\frac{\delta_{2}}{2}\cos\frac{\delta_{3}}{2}\\-\cos\frac{\delta_{1}}{2}\sin\frac{\delta_{2}}{2}\sin\frac{\delta_{3}}{2}\cos2[\alpha_{2}(\phi)-\alpha_{3}(\phi)]\\ -\sin\frac{\delta_{1}}{2}\cos\frac{\delta_{2}}{2}\sin\frac{\delta_{3}}{2}\cos2[\delta_{1}(\phi)-\delta_{3}(\phi)]\\-\sin\frac{\delta_{1}}{2}\sin\frac{\delta_{2}}{2}\cos\frac{\delta_{3}}{2}\cos2[\alpha_{1}(\phi)-\alpha_{2}(\phi)],
    \label{m11}
\end{multline}
\begin{multline}
\mathbbmss{B}_{q} = \sin\frac{\delta_{1}}{2}\cos\frac{\delta_{2}}{2}\cos\frac{\delta_{3}}{2}\cos2\alpha_{1}(\phi)\\ +\cos\frac{\delta_{1}}{2}\sin\frac{\delta_{2}}{2}\cos\frac{\delta_{3}}{2}\cos2\alpha_{2}(\phi)\\+\cos\frac{\delta_{1}}{2}\cos\frac{\delta_{2}}{2}\sin\frac{\delta_{3}}{2}\cos2\alpha_{3}(\phi)\\-\sin\frac{\delta_{1}}{2}\sin\frac{\delta_{2}}{2}\sin\frac{\delta_{3}}{2}\cos2[\alpha_{1}(\phi)-\alpha_{2}(\phi)+\alpha_{3}(\phi)],
\end{multline}
\begin{multline}
    \mathbbmss{C}_{q} = -\cos\frac{\delta_{1}}{2}\sin\frac{\delta_{2}}{2}\sin\frac{\delta_{3}}{2}\sin2[\alpha_{2}(\phi)-\alpha_{3}(\phi)]\\-\sin\frac{\delta_{1}}{2}\cos\frac{\delta_{2}}{2}\sin\frac{\delta_{3}}{2}\sin2[\alpha_{1}(\phi)-\alpha_{3}(\phi)]\\-\sin\frac{\delta_{1}}{2}\sin\frac{\delta_{2}}{2}\cos\frac{\delta_{3}}{2}\sin2[\alpha_{1}(\phi)-\alpha_{2}(\phi)],
    \label{m12}
\end{multline}
\begin{multline}
    \mathbbmss{D}_{q}= \sin\frac{\delta_{1}}{2}\cos\frac{\delta_{2}}{2}\cos\frac{\delta_{3}}{2}\sin2\alpha_{1}(\phi)\\+\cos\frac{\delta_{1}}{2}\sin\frac{\delta_{2}}{2}\cos\frac{\delta_{3}}{2}\sin2\alpha_{2}(\phi)\\+\cos\frac{\delta_{1}}{2}\cos\frac{\delta_{2}}{2}\sin\frac{\delta_{3}}{2}\sin2\alpha_{3}(\phi)\\-\sin\frac{\delta_{1}}{2}\sin\frac{\delta_{2}}{2}\sin\frac{\delta_{3}}{2}\sin2[\alpha_{1}(\phi)-\alpha_{2}(\phi)+\alpha_{3}(\phi)].
\end{multline}
It is evident that the real component of the off-diagonal elements is not equal to zero i.e., $\sigma_{q}\neq 0$. Consequently, this configuration cannot be considered equivalent to a single effective $q$-plate unless the term $\mathbbmss{C}_{q}$ is nullified. Furthermore, to represent an array of three $q$-plates as a single effective waveplate, two additional criteria must be fulfilled:
\begin{itemize} 
\item The fast axis orientation of the effective $q$-plate should lie within the plane of the plate and must exhibit an azimuthal dependence analogous to that of an individual $q$-plate. 
\item The effective $q$-plate must maintain uniform spatial retardance throughout its transverse plane.
\end{itemize}
The general analytical solution for the term $\mathbbmss{C}_{q}=0$ is mathematically intricate; therefore, in this work, we focus our analysis specifically on setups constructed from combinations of $q^{Q}$- and $q^{H}$-plates, yielding a total of eight distinct configurations.\\
\indent
Table \ref{table1} summarizes the conditions under which each of the eight possible combinations of three $q$-plates can effectively behave as a single equivalent $q$-plate. The table also presents the corresponding effective retardance $\delta_e$ and effective fast-axis orientation $\alpha_e$ for each configuration. From the table, it is evident that certain combinations yield behavior analogous to a $q^{Q}$-plate ($\delta_e = \pi/2$) and a $q^{H}$-plate ($\delta_e = \pi$). However, the cases labeled $\mathcal{E}_1$, $\mathcal{F}_1$, $\mathcal{G}_1$ and $\mathcal{H}_1$ exhibit a retardance that varies with the azimuthal angle $\phi$, indicating that the retardance is not uniform across the transverse plane of the resulting effective $q$-plate. This spatial non-uniformity violates the criteria for a well-defined effective waveplate. Consider the effective retardance for the $\mathcal{F}_1$ case ($q^{Q}q^{H}q^{Q}$ configuration), corresponding to the $q^{Q}q^{H}q^{Q}$ configuration, where the first and third $q$-plates have identical topological parameters, which is expressed as
\begin{equation}
\delta_{e} = \pi + 2\sin^{-1}\big[\cos2[(q_{3}-q_{2})\phi+(\alpha_{03}-\alpha_{02})]\big].
\end{equation}
To ensure that $\delta_e$ is independent of the azimuthal angle $\phi$, the topological charge of the plates must be equal, i.e., $q_3 = q_2$ $(=q_1=q)$. Under this condition, the expression for the effective retardance simplifies to
\begin{align}
\delta_{e} 
&= \pi + 2\sin^{-1}\left[\cos 2(\alpha_{03} - \alpha_{02})\right] \notag \\
&= 2\pi - 4(\alpha_{Q}-\alpha_{H})=2(\pi - 2\Delta\alpha).
\end{align}
In the above expression we have defined $\alpha_{03}=\alpha_{Q}$ as the offset angle of the $q^{Q}$-plate and $\alpha_{02}=\alpha_{H}$ as that of the $q^{H}$-plate and $\Delta\alpha=\alpha_{Q}-\alpha_{H}$. Hence, for case $\mathcal{F}_1$, the effective retardance can be tuned by controlling the offset angle. Fig. \ref{plot} (black curve) showing the variation of the effective retardance with the relative offset angle $\Delta\alpha = \alpha_{Q} - \alpha_{H}$. It is evident from the plot that by changing 
\begin{table*}[t]
\centering
\renewcommand{\arraystretch}{1.98}
\setlength{\tabcolsep}{12pt}  
\begin{tabular}{|c|c|c|c|c|}
\hline
\textbf{Setup}  & \textbf{Condition: $\mathbbmss{C}_{q}=0$} & \textbf{Cases} & $\delta_{e}$ & $\alpha_{e}=q_{e}\phi+\alpha_{0e}$ \\ \hline
$q^{H}q^{H}q^{H}$ & any $\alpha_{1}(\phi)$, $\alpha_{2}(\phi)$, $\alpha_{3}(\phi)$ & $\mathcal{A}_1$ & $\pi$ & $\alpha_{1}(\phi)-\alpha_{2}(\phi)+\alpha_{3}(\phi)$  \\ \hline
\multirow{2}{*}{$q^{H}q^{H}q^{Q}$}  & $\alpha_{1}(\phi)=\alpha_{2}(\phi)$ & $\mathcal{B}_1$ & \multirow{2}{*}{$\pi/2$} & \multirow{2}{*}{$\alpha_{3}(\phi)$} \\ 
                           & $|\alpha_{1}(\phi)-\alpha_{2}(\phi)| =\pi/2$ & $\mathcal{B}_2$ &  & \\ \hline
\multirow{2}{*}{$q^{H}q^{Q}q^{H}$}  & $\alpha_{1}(\phi)=\alpha_{3}(\phi)$ & $\mathcal{C}_{1}$ & \multirow{2}{*}{$\pi/2$} & \multirow{2}{*}{$2\alpha_{1}(\phi)-\alpha_{2}(\phi)$} \\ 
                           & $|\alpha_{1}(\phi)-\alpha_{3}(\phi)| =\pi/2$ & $\mathcal{C}_{2}$ &  & \\ \hline
\multirow{2}{*}{$q^{Q}q^{H}q^{H}$}  & $\alpha_{2}(\phi)=\alpha_{3}(\phi)$ & $\mathcal{D}_{1}$ & \multirow{2}{*}{$\pi/2$} & \multirow{2}{*}{$\alpha_{1}(\phi)$} \\ 
                           & $|\alpha_{2}(\phi)-\alpha_{3}(\phi)| =\pi/2$ & $\mathcal{D}_2$ &  &  \\ \hline
\multirow{2}{*}{$\textcolor{red}{q^{Q}q^{Q}q^{H}}$} & $2\alpha_{3}(\phi)=\alpha_{1}(\phi)+\alpha_{2}(\phi)$ & $\mathcal{E}_1$ & $\pi+2\sin^{-1}[\cos2\alpha_{32}(\phi)]$ & $-\frac{1}{2}\tan^{-1}[\cot2\alpha_{1}(\phi)]$ \\ 
                          & $|\alpha_{1}(\phi)-\alpha_{2}(\phi)|=\pi/2$ & $\mathcal{E}_2$ & $\pi$ & $\alpha_{3}(\phi)$ \\ \hline
\multirow{2}{*}{$\textcolor{red}{q^{Q}q^{H}q^{Q}}$}  & $\alpha_{1}(\phi)=\alpha_{3}(\phi)$ & $\mathcal{F}_1$ & $\pi+2\sin^{-1}[\cos2\alpha_{32}(\phi)]$ & $-\frac{1}{2}\tan^{-1}[\cot2\alpha_{3}(\phi)]$ \\ 
                           & $|\alpha_{13}^{+}(\phi)-2\alpha_{2}|=\pi/2$ & $\mathcal{F}_2$ & $\pi$ & $\alpha_{2}(\phi)$ \\ \hline
\multirow{2}{*}{$\textcolor{red}{q^{H}q^{Q}q^{Q}}$}  & $\alpha_{23}^{+}(\phi)=2\alpha_{1}(\phi)$ & $\mathcal{G}_1$ & $\pi+2\sin^{-1}[\cos2\alpha_{21}(\phi)]$ & $-\frac{1}{2}\tan^{-1}[\cot2\alpha_{3}(\phi)]$ \\ 
                           & $|\alpha_{2}(\phi)-\alpha_{3}(\phi)|=\pi/2$ & $\mathcal{G}_2$ & $\pi$ & $\alpha_{1}(\phi)$ \\ \hline 
\multirow{3}{*}{$q^{Q}q^{Q}q^{Q}$}  & $\alpha_{1}(\phi)=\alpha_{3}(\phi)$ & $\mathcal{H}_1$ & $2\cos^{-1}\left[-\frac{1}{\sqrt{2}}\cos2\alpha_{23}(\phi)\right]$ & $\mathcal{H}_{1}^{*}$ \\ 
                          & $|\alpha_{1}(\phi)-\alpha_{2}(\phi)|=\pi/2$ & $\mathcal{H}_2$ & $\pi/2$ & $\alpha_{3}(\phi)$ \\
                          & $|\alpha_{2}(\phi)-\alpha_{3}(\phi)|=\pi/2$ & $\mathcal{H}_3$ & $\pi/2$ & $\alpha_{1}(\phi)$ \\ \hline 
\end{tabular}
\caption{Realization of effective $q$-plate through the combination of three $q$-plates, each chosen as either a $q^{Q}$-plate or a $q^{H}$-plate. Here, the condtion $\sigma_{q}=0$ is the necessary for the configurations in order to be act as an effective $q$-plate. Notations are defined as: $\alpha_{x}=q_{x}\phi+\alpha_{0x}$, $\alpha_{xy}=\alpha_{x}-\alpha_{y}$ and $\alpha_{xy}^{+}=\alpha_{x}+\alpha_{y}$.}
\label{table1}
\end{table*}
$\Delta\alpha$ from $\pi/2$ to $0$ and from $3\pi/2$ to $\pi$, a continuous tuning of the effective retardance from $0$ to $2\pi$ is achievable. Additionally, by changing $\Delta\alpha$ from $\pi/2$ to $\pi$ and from $3\pi/2$ to $2\pi$, a continuous tuning of the effective retardance from $0$ to $2\pi$ is also achievable. Next, for the same case, the effective fast axis orientation is given by
\begin{align}
\alpha_{e}(\phi) 
&= -\frac{1}{2}\tan^{-1}\left[\cot 2\alpha_{3}(\phi)\right] \notag \\
&= q\phi + \underbrace{\left(\alpha_{Q} - \pi/4\right)}_{\substack{\text{Effective offset} \\ \text{angle}}}.
\end{align}
This implies that the effective topological charge of the combination $q^{Q}q^{H}q^{Q}$ is $q_{e}=q$ and the effective offset angle $\alpha_{0e}=\alpha_{Q}-\pi/4$. The above equation respects the definition of the fast axis orientation of the effective $q$-plate.\\
\indent
The configuration $q^{Q}q^{H}q^{Q}$ offers a distinct advantage by enabling a continuously tunable retardance spanning the full range from $0$ to $2\pi$, governed by the relative offset angles of the constituent plates. Moreover, it can be readily demonstrated that the configurations $q^{Q}q^{Q}q^{H}$ and $q^{H}q^{Q}q^{Q}$ similarly support tunable retardance given by $\delta_{e} = 2(\pi - 2\Delta\alpha)$, with the effective fast axis orientation given by 
$\alpha_{e}(\phi) = q\phi + (\alpha_{Q} - \pi/4)$. Therefore, the combination of two $q^{Q}$-plates and one $q^{H}$-plate, regardless of their sequential arrangement, demonstrates the tunable feature of the effective retardance under the concept of an effective waveplate.\\
\indent
The (effective) retardance corresponds directly to the rotation angle on the associated topological sphere \cite{umar20252}. Therefore, a configuration within a given index space that provides tunable retardance over the full range from $0$ to $2\pi$ is capable of realizing a complete $2\pi$ rotation on the associated topological sphere. In particular, the setups $q^{Q}q^{Q}q^{H}$, $q^{Q}q^{H}q^{Q}$ and $q^{H}q^{Q}q^{Q}$ exhibit tunable retardance if, and only if, all three $q$-plates share the same topological charge. This condition ensures that the setups belong to the index $q$-space. According to the definition of a topological index space \cite{umar2025holonomically}, the associated topological sphere is then the $q=\eta$ order sphere, thereby allowing these configurations to realize holonomic transformations on the corresponding sphere. As a result, such setups with topological charge $q$ are capable of performing a complete $2\pi$ rotation on the HOPS, showing the possibility of a full SU(2) walk on the HOPS. In our findings, the effective offset angle depends on $\alpha_{Q}$, therefore, by varying $\alpha_{Q}$, one can control the orientation of the rotation axis within the equatorial plane of the HOPS. Furthermore, by adjusting the relative offset angle $\alpha_{Q} - \alpha_{H}$, the desired amount of rotation on the sphere can be achieved. This result forms an important outcome of our work.\\
\indent
For all eight configurations, the effective fast axis orientation adheres to the relation $\alpha_{e}(\phi) = q_{e}\phi + \alpha_{0e}$, thereby satisfying one of the essential conditions to define an effective $q$ plate, as is evident from Table~\ref{table1}. For the $\mathcal{H}_{1}$ case ($q^{Q}q^{Q}q^{Q}$ configuration), the effective retardance is again a function of $\phi$, violating the requirement to be an effective $q$-plate. Therefore, taking $q_2=q_3$, we get
\begin{equation}
    \delta_{e} = 2\left[\pi-\cos^{-1}\left\{\frac{1}{\sqrt{2}}\cos2(\alpha_{02}-\alpha_{03})\right\}\right].
    \label{qqq}
\end{equation}
Here, the first and third plates are identical in each manner, but the middle plate is different in the context of the offset angle. This configuration also exhibits tunable retardance, though the range does not span the full interval from $0$ to $2\pi$. Instead, the retardance lying in the range $\pi/2$ to $3\pi/2$ and this can also be seen in Fig. \ref{plot}. Here, changing the $\Delta\alpha$ from $\pi/2$ to $0$, $3\pi/2$ to $\pi$, $\pi/2$ to $\pi$ and from $3\pi/2$ to $2\pi$, a continuous tuning of the effective retardance from $\pi/2$ to $3\pi/2$ is achievable. For the $\mathcal{H}_{1}$ case, the effective fast axis orientation is
\begin{widetext}
\begin{equation}
    \alpha_{e}(\phi)=\frac{1}{2} \tan^{-1}\left[\frac{\sin2\alpha_{2}(\phi) + \sin2[\alpha_{2}(\phi) - 2\alpha_{3}(\phi)] + 2\sin2\alpha_{3}(\phi)}{\cos2\alpha_{2}(\phi) - \cos2[\alpha_{2}(\phi) - 2\alpha_{3}(\phi)] + 2\cos2\alpha_{3}(\phi)}\right].
\end{equation}
\end{widetext}
This effective fast-axis orientation is also consistent with the conventional definition of the fast axis of the $q$-plate. It should be noted that the above expression should be dealt with under the condition $q_{2}=q_{3}$, as this ensures that the retardance of the configuration is free from azimuthal dependence.\\
\indent
The study of the effective waveplate concept for many combinations is carried out within a general framework, wherein the three constituent $q$-plates may possess arbitrary topological charges. An exception arises for configurations that exhibit tunable retardance: in order for such configurations to function as a single equivalent $q$-plate, the topological charge $q$ of all involved plates must be identical. Furthermore, those arrangements that allow arbitrary values of topological charge $q$ do not belong to the conventional topological index space but instead reside within a mixed index space.\\
\indent
An illustrative case of the mixed index space is the configuration of three $q^{H}$-plates (first configuration of Table~\ref{table1}). For this case, the effective retardance is $\delta_{e}=\pi$, and the effective fast axis is given by
\begin{equation}
\alpha_{e} = (q_{1} - q_{2} + q_{3})\phi + \alpha_{01} - \alpha_{02} + \alpha_{03}.
\end{equation}
Setting $q_{2}=0$ reduces the middle plate to a HWP with fast axis orientation $\alpha_{02}$, sandwiched between two $q^{H}$-plates of charges $q_{1}$ and $q_{3}$, gives
\begin{equation}
\alpha_{e} = (q_{1} + q_{3})\phi + \alpha_{01} - \alpha_{02} + \alpha_{03},
\label{mh}
\end{equation}
which gives the addition of the topological charges of the $q$-plates. Alternatively, taking $q_{3}=0$ (third plate as a HWP with fast axis orientation $\alpha_{03}$) leads to
\begin{equation}
\alpha_{e} = (q_{1} - q_{2})\phi + \alpha_{01} - \alpha_{02} + \alpha_{03},
\label{sh}
\end{equation}
representing the subtraction of the topological charge of the involved $q$-plates. Both results belong to the mixed index space and are fully consistent with the arithmetic with $q$-plates reported in \cite{delaney2017arithmetic}. From Eqs. (\ref{mh}) and (\ref{sh}), it is evident that the effective offset angle of the resultant $q$-plate is determined by the fast-axis orientation of the HWP. Therefore, by mechanically rotating the HWP within the setup, the offset angle of the resultant configuration can be tuned. This capability is particularly useful in applications where control over the offset angle is required.
\section{Conclusion}
\label{section04}
This paper presents a comprehensive mathematical formulation of the concept of the effective waveplate. We investigate how different configurations of three $q$-plates can function as a single effective $q$-plate, each of which is either a $q^{Q}$-type, a $q^{H}$-type, or a mixture thereof. A significant outcome of this study is that certain configurations behave as a single $q$-plate with fixed retardances of $\pi/2$ and $\pi$, whereas others, under specific conditions, exhibit tunable retardance spanning the range $0$ to $2\pi$. The tunable cases correspond to the combinations $q^{Q}q^{Q}q^{H}$, $q^{Q}q^{H}q^{Q}$, and $q^{H}q^{Q}q^{Q}$, with the tunability governed by the relative offset angle of the constituent plates. Moreover, the combination of three $q^{Q}$-plates results in a tunable retardance that ranges from $\pi/2$ to $3\pi/2$. A key result of this work is that the configurations supporting tunable retardance indicate the feasibility of achieving complete SU(2) coverage on the HOPS. Hence, this study may find a potential application for navigation on HOPS. A systematic investigation of full coverage on this topological sphere using such configurations naturally emerges as the next direction of our research. Moreover, this study may also find applications in advancing the controlled manipulation of structured light, particularly through precise polarization and phase transformations enabled by effective waveplate configurations.
\begin{acknowledgments}
\vspace{-0.1em}
\noindent
MU gratefully acknowledges the institute fellowship received from IIT Delhi, and also extends sincere thanks to the members of the Singular Optics Laboratory at IIT Delhi for their support and encouragement. PS acknowledges the financial support received from the Science and Engineering Research Board (SERB), India (CRG/2022/001267).
\end{acknowledgments}
\newpage
\nocite{*}
\bibliography{su2}
\end{document}